\documentclass[notitlepage, 10pt, floatfix, reprint, longbibliography, prl]{revtex4-2}
\usepackage[utf8]{inputenc}
\usepackage{amsmath}
\usepackage{amsfonts}
\usepackage{amssymb}
\usepackage{mathrsfs}
\usepackage{subfigure}
\usepackage[hidelinks]{hyperref}
\usepackage{outlines}
\usepackage{tikz}
\usepackage{booktabs}

\newcommand{\br}{\mathbf{r}}

\newcommand{\D}{\mathrm{d}}

\begin{document}

\title{Ion Mix Can Invert Centrifugal Confinement}
\date{February 24, 2026}
\author{E. J. Kolmes}
\email[Electronic mail: ]{ekolmes@princeton.edu}
\affiliation{Department of Astrophysical Sciences, Princeton University, Princeton, New Jersey 08540, USA}
\author{I. E. Ochs}
\affiliation{Department of Astrophysical Sciences, Princeton University, Princeton, New Jersey 08540, USA}
\author{N. J. Fisch}
\affiliation{Department of Astrophysical Sciences, Princeton University, Princeton, New Jersey 08540, USA}

\begin{abstract}
Centrifugal plasma traps, in which plasma is confined partly by centrifugal forces, represent a possible path to fusion energy production. 
In centrifugal plasma traps, electric fields naturally arise in the direction parallel to the magnetic field in order to ensure quasineutrality. 
These electric fields are sensitive to the particular mix of ion species present. 
This paper uses analytic and numerical calculations to consider how these mix effects could be used to improve device performance. 
Changing the composition of the plasma can mix or demix different ion species, improve confinement of certain ions, and can even expel certain species from the trap altogether. 
This makes it possible to generate an ``inverted centrifugal" end-plug with very promising properties. 
\end{abstract}

\maketitle

\textit{Introduction.} 
In plasma traps of various kinds, electric fields arise to enforce quasineutrality. 
These fields are sometimes called ambipolar fields. 
In magnetized plasmas, ambipolar fields are oriented parallel to the magnetic field lines. 
They are particularly important in linear devices like magnetic mirrors and centrifugal traps \cite{Pastukhov1974, Post1987}. 
Linear traps have been the subject of increasing interest in recent years, both for nuclear fusion and for other applications (such as nuclear waste processing and mass separation) \cite{Fowler2017, Yakovlev2018, Zweben2018, Inzhevatkina2021, Endrizzi2023, Schwartz2024, Beklemishev2024, Frank2025}. 
%

There have been a number of proposals to improve the performance of mirror-type configurations by manipulating the configuration to affect these fields. 
One such concept is the tandem mirror \cite{Baldwin1979, Tamano1995, Dimov2005, Fowler2017}, in which a central cell is placed between end-plug cells, with plasma density and temperature in the end-plugs chosen to engineer favorable ambipolar fields for the confinement of the central plasma. 
Another is the ``sloshing'' ion distribution, in which the ambipolar field is modified by a population of high-energy ions with a coherent turning point \cite{Kesner1973, Simonen1983, Post1987, Endrizzi2023}.

These ideas were developed for conventional (that is, non-centrifugal) mirrors. 
For centrifugal mirrors, the possibilities associated with these ambipolar fields have been relatively understudied. 
Centrifugal traps -- sometimes also called centrifugal mirrors or rotating mirrors -- achieve parallel confinement using centrifugal forces. 
If the magnetic field lines are arranged so that they bend radially inward near the ends of the trap, the centrifugal forces produce a confining effect along the field lines. 
Rotation can be driven by superimposing approximately-axial magnetic fields with approximately-radial electric fields; the electric fields can be driven by electrodes, neutral beams, or wave-particle interactions \cite{Fisch1992, Ellis2001, Fetterman2010, Gueroult2019, Rax2023, Beklemishev2024}. 
In this paper, we will consider the effects of ion species mix on ambipolar potentials in centrifugal traps. 
These mix effects lead to surprising outcomes. 
Adding more of one species can confine or deconfine other populations, or can lead to stratification effects in which different species are localized at different locations along field lines. 
Even more surprising is that mix effects can also be used to establish a highly advantageous centrifugal end-plug, analogous to the end-plugs used in conventional mirrors but relying on a different set of physical mechanisms. 
These effects suggest exciting new possibilities for centrifugal mirror fusion, among other applications. 

\textit{Simple model.} 
Consider a plasma centrifugal-mirror system containing electrons and some number of ion species. 
Consider the simple approximation in which the longitudinal confinement is entirely due to (1) the centrifugal potential and (2) the electrostatic potential. 
Suppose, furthermore, all species are Gibbs-distributed along each field line. 
We will revisit these assumptions later in the paper. 
Note, for now, that they mean neglecting the loss cone (that is, we take the centrifugal well to be substantially deeper than the species' thermal energies); they assume there are no other major nonthermal features; and they take the plasma to be non-relativistic. 
These assumptions also mean that we are considering the interior region of the trap; in the lower-density regions adjacent to where the magnetic field impinges on the walls, sheath and presheath electric fields are important, and may not conform to the model used here \cite{Beklemishev2024}. 

Keeping all of this in mind, for any species $s$, let the density along a given field line be given by
\begin{align}
n_s = n_{s0} \exp \bigg[ - \frac{q_s \Delta \phi}{T_s} - \frac{\mu_s \Delta \varphi_c}{T_s} \bigg] . \label{eqn:Gibbs}
\end{align}
Here $\Delta \phi$ is the electrical potential difference, $\Delta \varphi_c$ is the centrifugal potential energy as it would be experienced by a proton, $q_s$ is the charge, $\mu_s$ is the particle mass (normalized by the proton mass), and we have taken the temperature $T_s$ to be constant along the field line. 
$n_{s0}$ is a normalization constant. 
$\Delta \phi$ and $\Delta \varphi_c$ are both defined relative to some reference point on the field line. 
In the quasineutral limit, 
\begin{align}
n_e = \sum_i Z_i n_i, \label{eqn:quasineutrality}
\end{align}
where the sum over $i$ is over all ion species, the $e$ subscript refers to the electrons, and $Z_i \doteq q_i/e$ is the charge normalized by the elementary charge. 

\textit{Single-Bulk-Ion-Species Limit.}
If there is one bulk ion species for which $Z_i n_i$ greatly exceeds the charge-weighted densities of all other ion species, then 
\begin{align}
n_e \approx Z_i n_i
\end{align}
so that (taking the electron mass to be negligible)
\begin{align}
\frac{e \Delta \phi}{T_e} = - \frac{Z_i e \Delta \phi}{T_i} - \frac{\mu_i \Delta \varphi_c}{T_i} \, . \label{eqn:GibbsMatchInitial}
\end{align}
Then 
\begin{gather}
e \Delta \phi = - \bigg( \frac{T_e}{T_i + Z_i T_e} \bigg) \mu_i \Delta \varphi_c. \label{eqn:GibbsMatching}
\end{gather}
These expressions relate $\Delta \phi$ and $\Delta \varphi_c$ only along a given field line; cross-field variation of the potentials is important for other purposes \cite{Kolmes2024VoltageDrop} but does not come into the calculation here. 
Note that if $Z_i \gg 1$ and $T_i = T_e$, the relationship between $e \Delta \phi$ and $\Delta \varphi_c$ is set entirely by the charge-to-mass ratio. 

The single-bulk-species limit is already enough to see some of the implications of the self-consistent fields. 
Consider a plasma containing ion species $a$ and $b$. 
If $Z_a n_a \gg Z_b n_b$ (i.e., if $a$ is the bulk species), then the electric field is set by Eq.~(\ref{eqn:GibbsMatching}) with $i=a$ and the distribution along a field line is: 
\begin{gather}
n_a = n_{a0} \exp \bigg[ - \frac{1}{1 + Z_a (T_e/T_a)} \frac{\mu_a \Delta \varphi_c}{T_a} \bigg] . 
\end{gather}
In this limit, $\Delta \phi$ is set entirely by species $a$. 
A well in $\Delta \varphi_c$ always confines (attracts) species $a$ and a barrier in $\Delta \varphi_c$ always repels species $a$; the electrostatic field simply acts to reduce the efficacy of these wells (local minima) or barriers (maxima) by a factor of $1 + Z_a (T_e / T_a)$. 

If, on the other hand, $Z_a n_a \ll Z_b n_b$ (i.e., if $a$ is the trace species), then the electric field is set by Eq.~(\ref{eqn:GibbsMatching}) for $i=b$ and the distribution along a field line is: 
\begin{gather}
n_a = n_{a0} \exp \bigg[ - \bigg( \mu_a - \frac{Z_a T_e \mu_b}{T_b + Z_b T_e} \bigg) \frac{\Delta \varphi_c}{T_a} \bigg]. 
\end{gather}
The behavior of species $a$ depends dramatically on the quantity in parentheses. 
Species $a$ is pushed $\textit{away}$ from regions of lower $\Delta \varphi_c$ if 
\begin{gather}
\frac{T_e}{T_b + Z_b T_e} > \frac{\mu_a}{Z_a \mu_b} \, . 
\end{gather}
For example, if $T_e = T_b$, and if the bulk is fully ionized boron-11, the reversal condition becomes 
\begin{align}
\frac{\mu_a}{Z_a} < \frac{11}{6} \, . 
\end{align}
This implies, among other things, that an isothermal centrifugal well dominated everywhere by boron will simply not confine protons. 
The possibilities associated with this kind of reversal (whether for boron or for any other ions) are one of the main subjects of this paper. 

\begin{figure}
\centering
\includegraphics[width=0.91\linewidth]{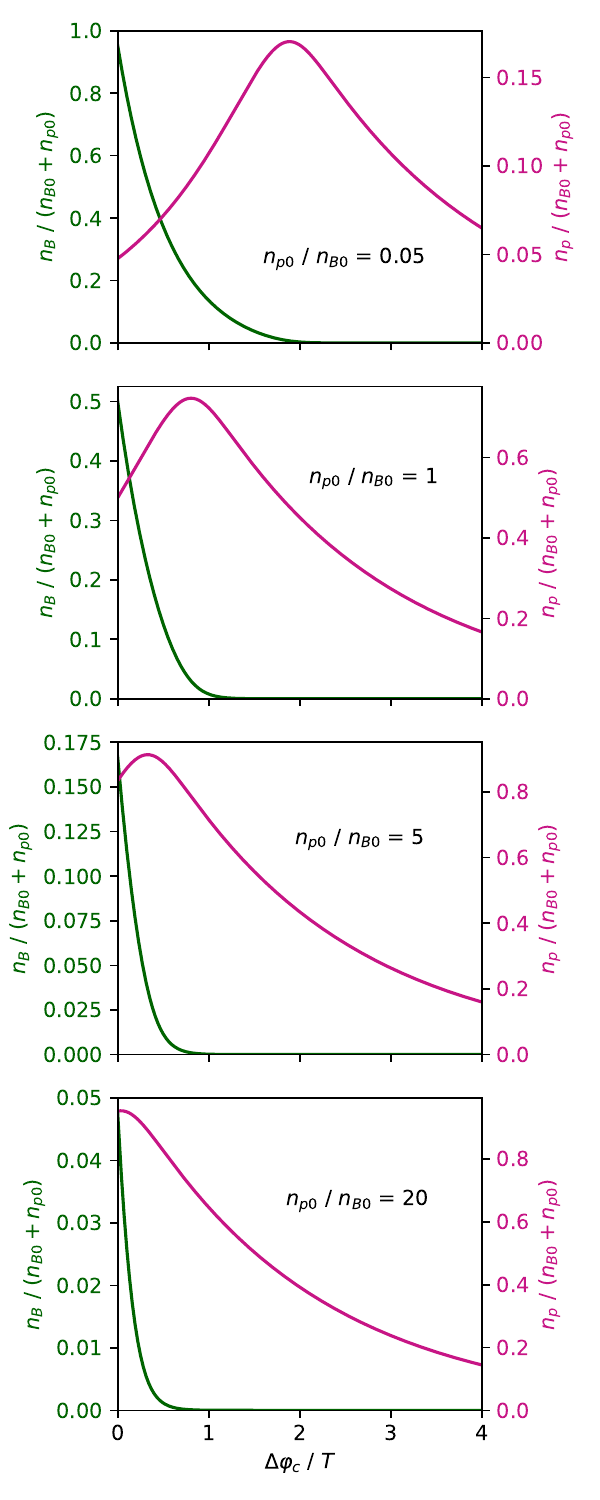}
\caption{Different distributions of protons and boron-11 along a field line, with field line position parameterized by centrifugal potential $\Delta \varphi_c$ (normalized by the temperature $T$). As the relative density of the protons drops, the self-consistent ambipolar fields push them further outwards (the top panel has the highest boron concentration; the bottom panel has the highest proton concentration). This example takes the ion and electron temperatures to be equal. Note that the overall normalization is arbitrary, so we could imagine the different cases taking place at fixed total ion charge or fixed total ion number. $n_{p0}$ and $n_{B0}$ are defined as the proton and boron densities at $\Delta \varphi_c = 0$.} \label{fig:pB11}
\end{figure}

\textit{More General Particle Mixes.}
Now consider the case in which several ion species contribute to the shape of the ambipolar potential. 
Take $n_{s0}$ to be fixed. 
Then the problem is to calculate the $\Delta \phi$ that solves 
\begin{align}
n_{e0} \exp \bigg[ \frac{e \Delta \phi}{T_e} \bigg] &= \sum_i Z_i n_{i0} \exp \bigg[ - \frac{Z_i e \Delta \phi}{T_i} - \frac{\mu_i \Delta \varphi_c}{T_i} \bigg] . \label{eqn:multiSpeciesGibbsBalance}
\end{align}
We can parameterize the shape of $\Delta \phi$ by $\Delta \varphi_c$. 
This may not be analytically tractable, but it is numerically straightforward. 

For example, Fig.~\ref{fig:pB11} shows the behavior of a plasma containing protons and boron-11 for different relative concentrations of the two ions. 
In the high-proton-fraction limit, both species sit at the bottom of the well. 
As more boron is added, the protons are pushed away. 
When parsing these results, it is helpful to keep in mind that this model has no notion of deconfinement; Eq.~(\ref{eqn:multiSpeciesGibbsBalance}) represents the limit of an infinitely deep potential well. 
In the high-boron-fraction limit, the protons cannot be pushed out of the well; instead, they bubble up in the potential well until they get to a height where the boron no longer dominates and they can be confined. 
Our intuition for a finite well should be that a species can eventually be pushed out, not just pushed to ever-higher potential. 

A more subtle feature of Fig.~\ref{fig:pB11} is that the boron is more tightly bound to the bottom of the well in scenarios where it is colocated with more protons. 
This is a general feature of these ambipolar fields: if one ion species tends to produce larger ambipolar electric fields than another, then higher concentrations of the first species (i.e., the species with higher $m_i/(Z_i+1)$) result in worse confinement for both, but higher concentrations of the second species result in \textit{better} confinement for both. 

\begin{figure}
\centering
\includegraphics[width=\linewidth]{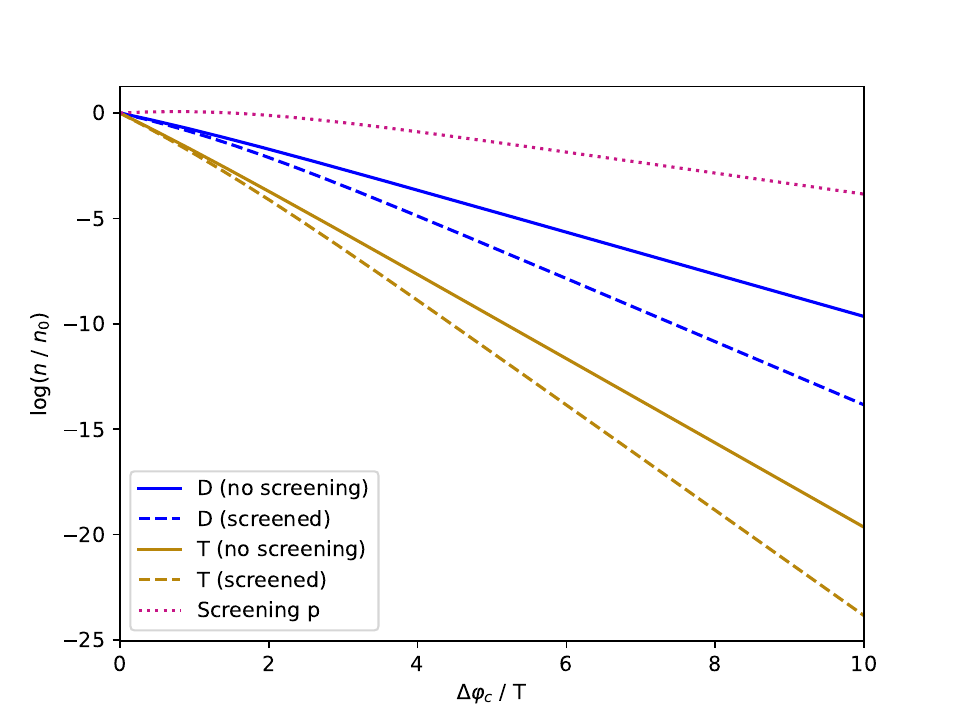}
\caption{The introduction of a small fraction of protons can improve the confinement of deuterium and tritium. The example without screening has 50\% deuterium and 50\% tritium; the example with screening has 10\% protons, 45\% deuterium, and 45\% tritium locally at the bottom of the well.} \label{fig:protonScreening}
\end{figure}

Consider, then, another example: a deuterium-tritium plasma that is ``doped" with a smaller fraction of protons. 
Protons, with their high charge-to-mass ratio, tend to reduce the ambipolar fields wherever they are found. 
This is illustrated in Fig.~\ref{fig:protonScreening}, where the introduction of a small population of protons substantially improves the confinement of deuterium and tritium in a centrifugal well. 
The example in the figure introduces a relatively small proton fraction; the effect on deuterium and tritium confinement depends on the proton fraction per Eqs.~(\ref{eqn:Gibbs}) and (\ref{eqn:quasineutrality}).  

So, should deuterium-tritium centrifugal traps operate with a small fraction of protons? 
The answer depends on which factors limit the performance of the trap. 
The introduction of protons allows the deuterium and tritium to be confined at lower rotation speeds, which would suggest that a pressure-limited device could operate at higher plasma densities (the limits on the plasma pressure are less stringent if the plasma rotates more slowly, since the magnetic field does not need to counteract such large centrifugal forces \cite{Bekhtenev1980}). 
Note that there will be some loss of fuel density due to dilution; in the limiting case where all of the fuel has been replaced by protons, the reactivity vanishes altogether. 
In favorable scenarios (particularly when the proton fraction is not too large) this can be more than offset by the increase in overall plasma density; see Fig.~\ref{fig:protonScreening}. 
However, the protons themselves are relatively poorly confined in these scenarios. 
This downside might be mitigated using the centrifugal end-plugs proposed later in this paper. 

The possibility of better parallel confinement addresses one of the major challenges facing centrifugal traps: strong centrifugal requirement requires fast rotation, and fast rotation requires imposing large electric fields across the magnetic flux surfaces. 
If the same confinement can be achieved with smaller electric fields, then all of the technical challenges associated with these large fields become less severe. 

The same physical mechanisms discussed in the two examples above can also be applied to a wide variety of other problems. 
One of these is the physics of heavy impurities. 
Because heavy impurities are often not fully ionized, their charge-to-mass ratio can be particularly low. 
This means that impurity accumulation has the potential to worsen the longitudinal confinement of all other species. 
Another straightforward application is plasma mass filters, which are designed to separate an input of mixed ions according to species \cite{Lehnert1971, Krishnan1981, Krishnan1983, Fetterman2011b, Gueroult2015, Dolgolenko2017, Zweben2018, Gueroult2018ii}. 
If the composition of a centrifugal trap can be tuned to confine or deconfine different species according to mass and charge, the configuration could be used for separations. 
Centrifugal potentials can also be important in toroidal devices like tokamaks. 
Distributions of different species along field lines in toroidal devices are often studied using the same analytic models employed here \cite{Wesson1997, Casson2010, Angioni2014Tungsten}, and many of the ideas suggested here could readily be applied to those systems. 

\textit{A Novel End-Plug.} 
Thus far, the discussion has focused on the behavior of ions sitting in a centrifugal well: mixing, demixing, confinement, and deconfinement. 
As we have seen, there are scenarios in which a centrifugal well can become repulsive to certain species, acting instead as a potential barrier. 
For a single isolated well, this looks like a failure of confinement. 
But one could just as well imagine using two or more of these barriers to better confine some central region -- much in the same way that repulsive end cells are used in conventional tandem mirrors. 

\begin{figure}
	\includegraphics[trim={13.5cm, .5cm, 0cm, 0cm},clip,width=\linewidth]{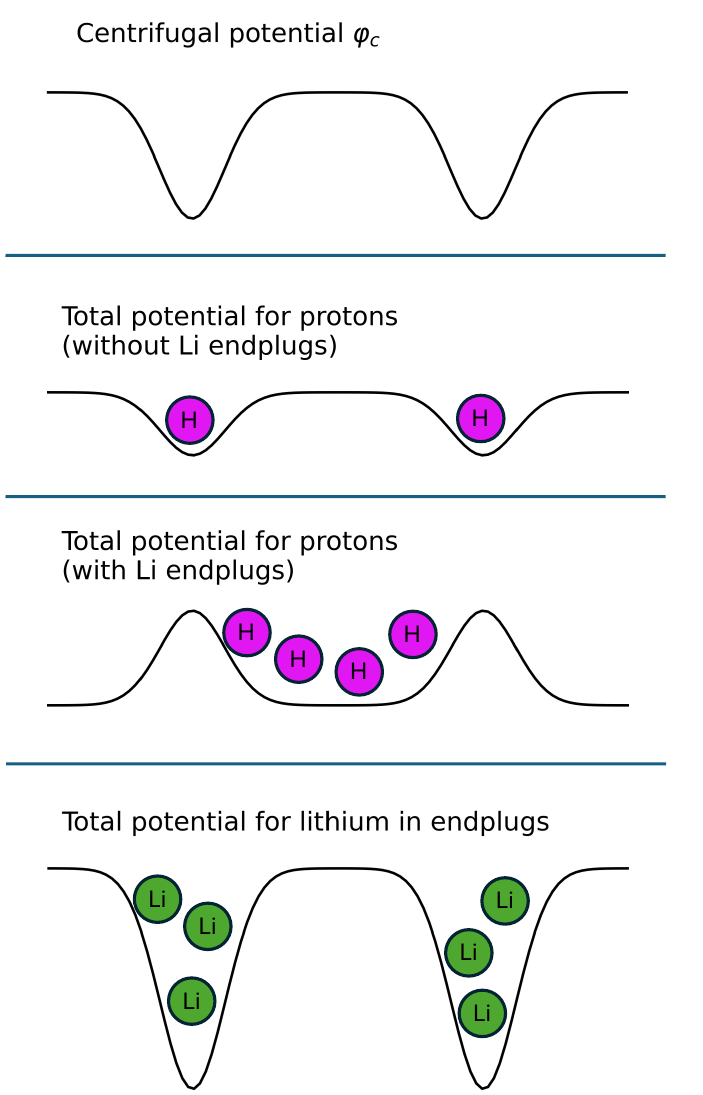}
	\caption{A cartoon of a centrifugal end-plug, in which a barrier species (in this case, lithium) inverts the centrifugal potential in two end regions in order to create a barrier for a central species (in this case, hydrogen). The top panel shows the centrifugal potential. 
	The second panel shows the total (centrifugal plus electrostatic) potential experienced by protons if only protons occupy the centrifugal wells.  
	The third and fourth panels show the total potentials experienced by protons and lithium if the wells are occupied by lithium. } 
	\label{fig:endplugCartoon}
\end{figure}

For example, consider a population of protons. 
For simplicity, take all species to have the same temperature. 
If the protons are placed (without other ions) in a simple centrifugal well with potential $\Delta \varphi_c$, they see a total confining potential 
\begin{gather}
\Delta \varphi_c + e \Delta \phi = \frac{1}{2} \Delta \varphi_c. 
\end{gather}
This follows from Eq.~(\ref{eqn:GibbsMatching}). 
The confinement generally gets worse if other ions are placed in the well alongside the protons. 
However, suppose instead that the protons are placed in a region between two centrifugal wells, each populated with some other, lower charge-to-mass ratio ion -- say, $^{7}$Li. 
In the limit where the centrifugal wells are populated entirely with lithium, the protons are repelled from the wells, seeing a total potential  
\begin{gather}
\Delta \varphi_c + e \Delta \phi = - \frac{3}{4} \Delta \varphi_c. 
\end{gather}
In other words, counter-intuitively, for a given well depth $\Delta \varphi_c$, it is more efficient for the purposes of proton confinement to set up a \textit{barrier} full of some other, lower-$Z/m$ species than it is to use that same $\Delta \varphi_c$ as a well to directly trap the protons. 
This is illustrated in Figure~\ref{fig:endplugCartoon}; the centrifugal wells in the figure correspond to the end-plugs, with the region between the wells corresponding to the central cell \footnote{See Supplemental Material [URL] for a more quantitative scenario.}.
In order to easily maintain the different compositions in the different segments of the trap, it would be important to make sure that each species is sufficiently well-trapped within its intended region (as in the case in the figure). 
In some cases, it might be helpful to ensure that the potential barrier between end-cell and the exterior of the trap is weaker than that between end-cell and central cell (for the end-cell species), so that end-cell ions diffuse out of the trap before entering the central cell. 
One could make an even more effective trap by placing the protons in a centrifugal well of their own, then flanking that well with end-cells full of another species. 

This mechanism is particularly well-suited to trapping protons because of their high charge-to-mass ratio. 
For other ions, it is often still possible to find another species with a lower charge-to-mass ratio to use in an endplug. 
However, there are also other ways to use the same mechanism to trap relatively low-$Z/m$ species. 

One possibility is to use proton ``doping" (as in the example in Fig.~\ref{fig:protonScreening}) to improve the confinement of another species in a well, then to use end-cells to prevent the protons themselves from becoming a serious loss channel. 
This strategy appears to be quite broadly applicable, so long as the application can tolerate the presence of a small population of protons. 

Another possibility is to use partially ionized species in the end-cell. 
For partially ionized species, the charge-to-mass ratio can be very small, so an end-cell populated by a partially ionized species can in principle produce a very strong barrier. 
For higher-temperature applications, there is the challenge of how to maintain a lower ionization state (and the radiative losses normally associated with partially ionized species like tungsten in high-temperature plasmas). 
A degree of thermal isolation between the main cell and the end-cells naturally arises if the end-cells produce potential barriers to repel main-cell ions, but the limits on how easily a temperature difference might be maintained are left as an open question. 
The ``thermal barrier" concept for the end-cells of conventional (non-centrifugal) mirrors does rely on maintaining a temperature difference between end-cell and main-cell, and least for the electron population \cite{Baldwin1979, Post1987}. 
However, it is substantially easier to maintain a difference in electron temperatures than in ion temperatures. 
Of course, in the limit of very low barrier species charge-to-mass ratio, it would be important to maintain a large enough magnetic field in the end-cells to maintain perpendicular confinement. 

\textit{Discussion.} 
Surprising and unexplored phenomena occur in centrifugal traps with mixtures of ion species. 
Changing the ion composition can cause different species to mix or demix in different parts of the trap. 
For any given species, it can also improve or degrade the trap's confinement properties, even to the point that a centrifugal well can invert to become a barrier. 

Each of these effects opens up an array of possibilities in applications ranging from controlled fusion to mass filters. 
Our hope in this paper is not to fully explore these possibilities, but to identify a class of very curious and potentially useful effects in multispecies rotating plasma. 
%

The model used here comes with a number of assumptions. 
%
First, and perhaps most importantly: we take each species to be Gibbs-distributed along field lines. 
This assumption is examined in further detail in the Appendix. 
Kinetically, it is equivalent to the assumption that the velocity distribution at the midplane is Maxwellian. 
This requires that the plasma be deeply (centrifugally) trapped, so that loss regions do not significantly modify the kinetics. 
It also requires that no other major nonthermal features are present. 
This assumption would fail, for example, in the presence of a 	large ``sloshing" population of beam ions. 
However, it is consistent with experimental observations in the literature \cite{Teodorescu2010, Angioni2014Tungsten}. 
The same assumption also takes the plasma to be non-relativistic. 
At high temperatures (i.e., $T_e$ approaching 511 keV), relativistic effects become important \cite{Ochs2023AmbipolarPotentials}. 

Second, the assumption of quasineutrality is safe only in the core of the plasma. Near the boundaries of the plasma, there are sheath regions in which quasineutrality fails. 
This is true in both conventional mirrors \cite{Pastukhov1974, Cohen1978, Cohen1980, Najmabadi1984} and centrifugal traps \cite{Bekhtenev1980, Volosov1981, Beklemishev2024}. 
These sheaths are necessary in order to understand the relationship between the model used in this paper and the treatments of self-consistent electric fields found elsewhere in the literature. 
It is common to calculate the total voltage drop across the plasma by equating the ion and electron particle loss rates, the idea being that the plasma will charge positive until these rates balance. 
The sheaths explain how the total voltage drop required by matching loss rates can be consistent with the local voltage profile predicted by quasineutrality. 
A more detailed treatment of the sheaths could be desirable in order to apply these ideas to devices like fusion thrusters. 
For example, in a deuterium-tritium rocket, tritium may be much more valuable to retain than deuterium. 
The quasineutrality assumption means that much of the analysis presented here would not apply as written to non-neutral traps like those sometimes used to contain antimatter \cite{Dubin1999, Surko2004, Baker2008}. 
However, one can anticipate adapting many of the underlying ideas to those applications -- particularly the use of one species to generate fields that trap another species. 

How these results are modified in other cases is largely left for future work. 
However, some of the physics described here is more general than the model would suggest. 
Eqs.~(\ref{eqn:GibbsMatchInitial}) and (\ref{eqn:GibbsMatching}), and much of the following discussion, apply any time the density distribution of each species along a field line is a function of the Gibbs factor alone. 
This is also true, for example, of the truncated-Maxwellian loss-cone model discussed in the Appendix. 

The ideas presented here would also apply to forces other than centrifugal forces. 
Ponderomotive forces have been proposed for various uses in plasma traps \cite{Watari1978, Fader1981, Rubin2023}; these forces can have various mass and charge dependences, depending on the polarization of the fields. 
The same analysis would also apply to gravitational forces. 
Indeed, the mechanisms described here are closely related to effects that have been noted in the astrophysics literature in which protons can be expelled from stars, depending on the stars' compositions \cite{Montmerle1976, Michaud1979}. 
Other forces would appear in the Gibbs factor in Eq.~(\ref{eqn:Gibbs}), consistent with a zero-current force balance. 
However, the most immediate application is a new way of end-plugging centrifugal mirror machines. 
Centrifugal mirrors have advantages over their conventional counterparts, both in terms of confinement and in terms of stability \cite{Lehnert1971, Bekhtenev1980, Piterskii1995, Ellis2005, Teodorescu2010, Kolmes2024Flutes}. 
If the techniques proposed in this paper can further improve the performance of centrifugal mirror traps, the upside potential could be significant. 

\textit{Acknowledgments.} The authors are grateful to Alex Glasser, Mike Mlodik, and Tal Rubin for helpful conversations. 
This work was supported by ARPA-E Grant No. DE-AR0001554. 

\section*{Data Availability Statement}

The data that support the findings of this article are openly available at \cite{KolmesCentrifugalInversionPlotData}. 

\bibliography{../../../Master.bib}

\appendix 
\section{Appendix: More General Distributions}

\begin{figure}
	\centering
	\includegraphics[width=1\linewidth]{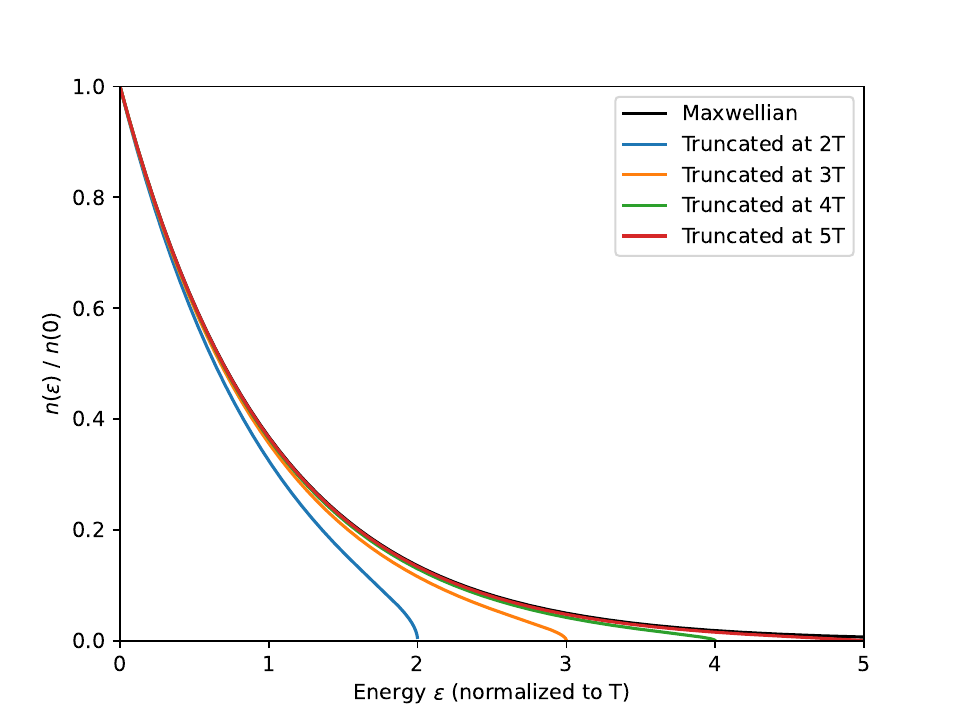}
	\caption{Spatial density distributions truncated at different levels, with space parameterized by the local potential energy (including centrifugal and electrostatic parts). }
	\label{fig:truncatedDistributions}
\end{figure}

The kinetic distribution of particles may not always be Maxwellian. 
Consider some more general distribution $f(\varepsilon, \mu)$ in constants-of-motion space, where $\varepsilon$ is the particle energy and $\mu$ is the magnetic moment. 
The spatial distribution along a field line with midplane distribution $f(\varepsilon, \mu)$ can be calculated as follows: 
\begin{align}
	n(\br) &= \int_0^\infty \D \mu \int_{B \mu + U}^\infty f(\varepsilon, \mu) \sqrt{g} \D \varepsilon \\
	&= \int_0^\infty \D \mu \int_{B \mu + U}^\infty \frac{4 \pi B f(\varepsilon, \mu) \D \varepsilon}{\sqrt{2(\varepsilon - \mu B - U)}} . 
\end{align}
Here $U$ is the total potential for the species at point $\br$ (including centrifugal and electrostatic parts); $\sqrt{g}$ is the usual Jacobian volume factor for the change of coordinates between constants of motion at the midplane and at $\br$; and $B$ is the local strength of the magnetic field. 

Consider the case where $B$ is constant along a field line. 
Practically, in a mirror-like configuration this would require internal shaping coils to achieve. 
Configurations with inner coils are of significant interest \cite{Bekhtenev1980, Volosov1981, Volosov2006}; for present purposes, they have the advantage of analytic simplicity. 
In this case, 
\begin{gather}
	n(\br) = \int_0^\infty \D \mu \int_U^\infty \frac{4 \pi B f(\varepsilon_{||},\mu) \D \varepsilon_{||}}{\sqrt{2 (\varepsilon_{||} - U)}} \, . 
\end{gather}
Here $\varepsilon_{||}$ is the parallel component of the energy. 
Let 
\begin{gather}
	g(\varepsilon_{||}) \doteq \int_0^\infty \D \mu \, B f(\varepsilon_{||}, \mu). 
\end{gather}
Then, taking $B = \text{const}$, 
\begin{gather}
	n(\br) = \frac{4 \pi}{\sqrt{2}} \int_U^\infty \frac{g(\varepsilon_{||}) \, \D \varepsilon_{||}}{\sqrt{\varepsilon_{||} - U}} \, . 
\end{gather}
Suppose particles are lost if they get past some maximal energetic barrier height $\varepsilon_\text{max}$. 
Different analytic models for loss-cone distributions in low-collisionality plasmas exist \cite{Li2025}, but one straightforward choices is to model the distribution as a ``truncated Maxwellian," with 
\begin{align}
	g(\varepsilon_{||}) = A e^{-\varepsilon_{||} / T} \Theta (\varepsilon_\text{max} - \varepsilon_{||}) ,
\end{align}
where $\Theta$ is the Heaviside step function. 
Then, taking $U < \varepsilon_\text{max}$, 
\begin{align}
	n(\br) &= \frac{4 \pi A}{\sqrt{2}} \int_U^{\varepsilon_\text{max}} \frac{e^{-\varepsilon_{||} / T} \, \D \varepsilon_{||}}{\sqrt{\varepsilon_{||} - U}} \\
	&= 2^{3/2} \pi^{3/2} A T^{1/2} e^{-U/T} \text{erf} \bigg[ \sqrt{\frac{\varepsilon_\text{max}-U}{T}} \bigg] . 
\end{align}
If $n(\br_0) = n_0$ at the midplane location $\br_0$, and if $U(\br_0) = U_0$, then (taking $T$ to be constant), 
\begin{gather}
	n(\br) = n_0 e^{-(U - U_0) / T} \, \frac{\text{erf} \sqrt{(\varepsilon_\text{max} - U) / T}}{\text{erf}\sqrt{(\varepsilon_\text{max}-U_0)/T}} \, .
\end{gather}
These density distributions are plotted in Figure~\ref{fig:truncatedDistributions}. 
The corrections to the density distribution relative to the untruncated Maxwellian are relatively mild for $\varepsilon_\text{max} \gtrsim 3$. 

\begin{figure}
	\centering
	\includegraphics[width=1\linewidth]{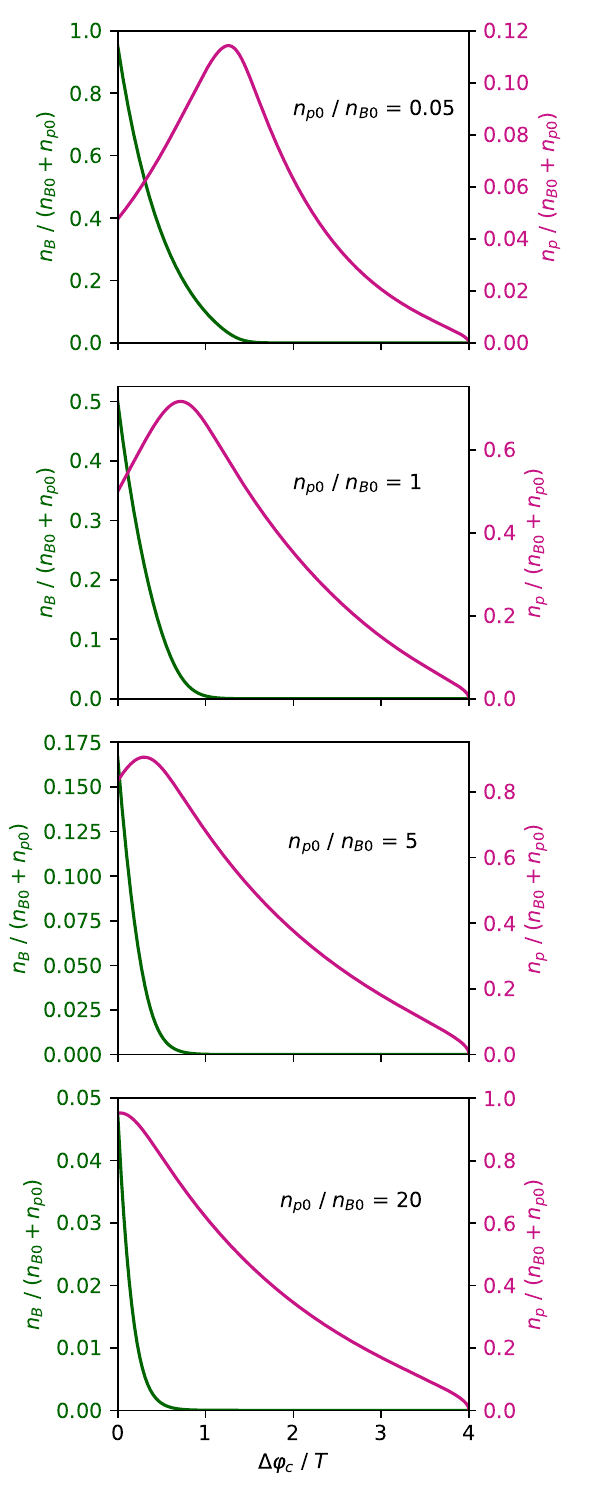}
	\caption{Different distributions of protons and boron-11 along a field line, much the same as is shown in Figure~\ref{fig:pB11}. However, where the distributions in Figure~\ref{fig:pB11} assume that the particles are Gibbs-distributed along field lines, the distributions used here instead assume truncated Maxwellian distributions, with a cutoff at $4T$. }
	\label{fig:pB11distributionsTrunc}
\end{figure}

In order to get a sense for the effects of truncation on the results discussed in this paper, we can redo the numerical calculation of proton and boron-11 distributions shown in Figure~\ref{fig:truncatedDistributions}. 
The version using truncated distributions is shown in Figure~\ref{fig:pB11distributionsTrunc}. 
In the cases shown in the figure, all species are truncated at $4T$. 
There is some visible distortion of the curves relative to their behavior without truncation, but the end results are qualitatively very similar. 

In principle, the particle loss conditions (and therefore the truncation thresholds) should be modified by the appearance of sheath or presheath fields at the edges of the device. 
The results shown in Figure~\ref{fig:pB11distributionsTrunc} are not the last word on the issue, but they suggest that the behavior of $n(\br)$ is reasonably robust against the effects of loss cones, so long as particles are reasonably well-trapped. 

This same formalism can also be used to illustrate cases in which nonthermal features do play a major role. 
Consider, for example, a beam population with midplane kinetic distribution 
\begin{gather}
	f_\text{beam}(\varepsilon, \mu) = \alpha \delta(\mu - \bar{\mu}) \delta(\varepsilon - \bar{\varepsilon})
\end{gather}
for some constants $\alpha$, $\bar \mu$, and $\bar \varepsilon$. 
The resulting spatial density distribution would be
\begin{gather}
	n_\text{beam} = \frac{4 \pi \alpha B}{\sqrt{2 (\bar \varepsilon - \bar \mu B - U)}} \, . 
\end{gather}
The spatial distribution is peaked because the particles have a coherent turning point. 
Quasineutrality would require a corresponding peak in the electrostatic potential. 
Features in the electrostatic potential that are driven by this kind of nonthermal structure need not correspond to any particular features in the centrifugal potential. 
Such features will tend to distinguish between ion species based on charge rather than charge-to-mass ratio, which raises a variety of interesting possibilities. 
We leave a more complete study of various nonthermal distributions and their implications to future work. 

Note that distributions of the form $f(\varepsilon, \mu)$ implicitly assume a low-collisionality regime where $\mu$-invariance is respected (e.g., over the course of many bounces). 
The analysis in the main body of this paper would apply whether the plasma is collisional or not. 
When we discuss non-constant temperatures, in the low-collisionality limit we typically have in mind multiple potential wells where the trapped populations in each well may have different characteristic energies. 
It is also worth pointing out that even in the ``collisionless'' limit (where the collision time is long compared with the bounce time), collisions may still encourage $f$ to be more nearly Maxwellian, if the collision time is short compared with the confinement time. 
This is part of the reason why it makes sense to focus on near-Maxwellian models. 

\end{document}